# Efficient Group Key Management Schemes for Multicast Dynamic Communication Systems

Muhammad Yasir Malik

2012

# Abstract

Key management in multicast dynamic groups, where users can leave or join at their ease is one of the most crucial and essential part of secure communication. Various efficient management strategies have been proposed during last decade that aim to decrease encryption costs and transmission overheads. In this report, two different types of key management schemes are proposed. First proposed scheme is based on One-way function tree (OFT). The proposed scheme fulfills the security gaps that have been pointed out in recent years. Second proposed scheme is based on logical key hierarchy (LKH). This proposed scheme provides better performance for, rather inflexible and expensive, LKH scheme.

**Keywords:** Group communication, key management (KM), logical key hierarchy (LKH), one-way function tree (OFT)

# Table of Contents

# List of Figures

# List of Tables

# 1 Introduction

## 1.1 Background

With increasing communication services, users are often grouped in various applications. They normally form centralized or decentralized structures, capable of handling entities involved in functions ranging from web and mail to sensor networks, file sharing to databases, and so on. Applications of such groups are enormous, and so are the demands for secure and reliable communication in these groups.

Internet protocol (IP) multicast, also known as multicast, is used to share contents with multiple users in a group. This form of communication tends to be efficient in terms of bandwidth as compared to unicast protocols, as it transmits information to every user in the group simultaneously. Internet group management protocol (IGMP) [9] is an example of multicast systems, in which any member can broadcast data to all $n$ members in the group. Any user can join and receive contents in IGMP, which makes it a scalable system.

Lack of access control and authentication poses security threats to IGMP, as any host can send and receive data from these systems. Conventional method to enforce restrictions on data flow is use of encryption to secure data contents, such that only the desired hosts can gather the data by utilizing cryptographic keys. Only the members with appropriate key can decrypt the data, which makes the communication secure and reliable.

Key distribution centre (KDC) or server is responsible for authentication of a user interested in joining the group. Server authenticates the user and allocates location to newly joined user. Server also provides the necessary keys to the user, which enables this user to communicate within the group.

Key management protocols are responsible for key pre-distribution and key updating in case of changes in the group. Group keys are shared among all the members and the contents to be shared are encrypted with this key and broadcasted to all group members. Such groups are supposed to be flexible enough to allow new hosts to join and present members to leave. Joining and leaving of hosts require change in the group key, so that the privacy and secrecy of the group members and their communication can be preserved.

Figure 1-1 shows structure of a secure centralized multicast system.

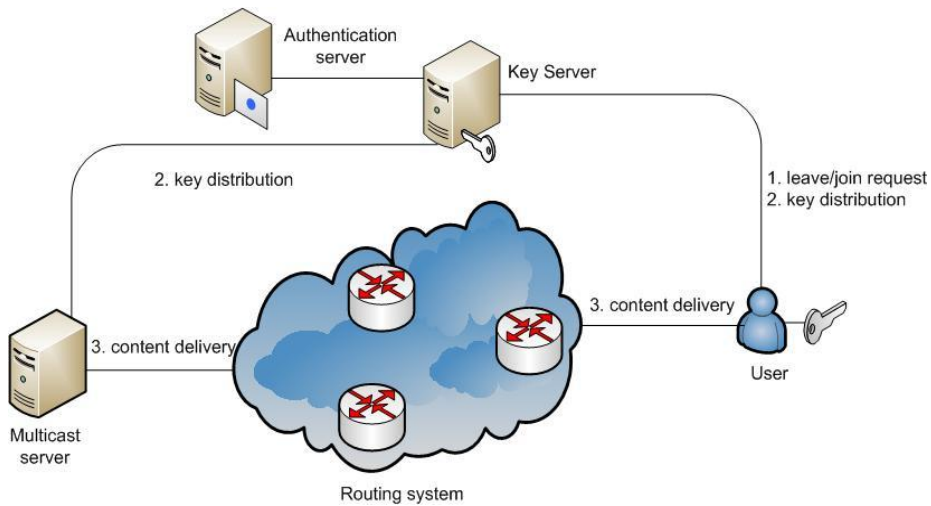

Figure 1-1 Secure centralized multicast system.

To maintain group keys, secure key management protocols are devised and employed. These protocols provide the authentication services, along with changing of the group keys with each user joining and leaving. The process of changing keys on every user join or leave is called *key updating* or *rekeying*.

Lack of presence of any key management protocol has rekeying cost of *nK* for *n* users. Logical key hierarchy (LKH) [5] and one-way function key tree (OFT) [1] are two much efficient centralized key management schemes. These both schemes differ in their functionalities; OFT follows down-up strategy as opposed to LKH which follows a static key tree structure. LKH has broadcast costs of $2hk + h$ for *n* users, where *k* is key size in bits and *h* is the height of tree. OFT scheme includes users along with the server in key updating process. This makes the scheme more efficient and it cuts the overhead cost at rekeying by $1/2$, i.e., $hk + h$.

## 1.2 Security Requirements

Centralized key management schemes must, in all conditions, fulfill some security requirements. Their basic security requirements are forward and backward secrecy in the group. Security requirements are discussed in detail in Chapter 2.

**Forward Secrecy:** Evicted members of the group are unable to access new information in the group, which states they cannot compute (or access) newer group keys.

**Backward Secrecy:** New members of the group are unable to access previous information in the group, which states that they cannot compute older group keys.

## 1.3 Objectives

To fill the security gap caused by collusion attacks on OFT and to reduce comparatively higher cost of LKH scheme, several improvements in both of these key management schemes have been proposed in this thesis.

OFT scheme is found to be weak against attacks by adversaries. We propose an improved OFT scheme, which guarantees better security at minimum costs. We also propose a simple LKH scheme for key distribution which provides the same functionalities at lower transmission cost.

Our improvements guarantee less cost in both schemes.

# 2 Key Management Schemes

This chapter describes two main centralized key management schemes in detail; logical key hierarchy and one-way function key tree. This chapter also covers collusion attacks on OFT scheme and revisited security requirements for key management schemes.

## 2.1 Logical Key Hierarchy

LKH maps all members of the group as leaves of a structured tree, most commonly as a balanced binary tree. Group key is at the root of the tree, whereas the leaves represent group members. Group members have to store their individual keys, group key, and all node keys in the path from member to the group key.

Figure 2-1 shows a balanced binary LKH key tree with height $h = 3$. If the number of users in the group is $n$, then height of group is given as $h = \log_2 n$. On each user leave or join, group key and other node keys in the path must be changed. New group key can be distributed by following the algorithms for join and leave functions, which will be defined in next subsections.

The complexity for key distribution to $n$ users will be $O(\log n)$.

### 2.1.1 LKH Tree Structure

All users in the group store $\log_2 n + 1$ keys, out of which one is their individual key and all other $h$ keys belong to the middle nodes in its path to the group node key. These $h$ keys with the user need to be changed after each join or leave.

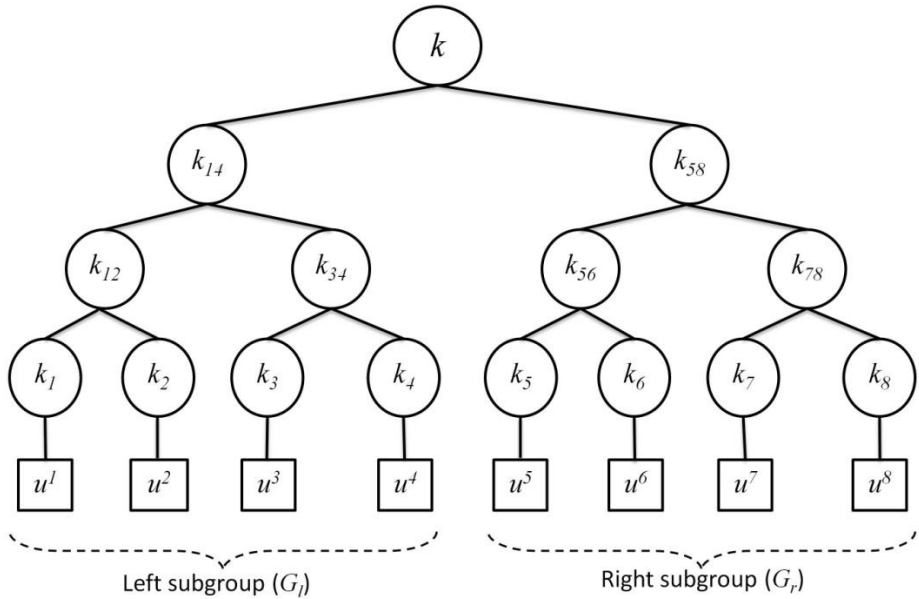

Figure 2-1 Logical key tree.

### 2.1.2 On User Join

Suppose user $u^8$ joins the group and users $u^1$ to $u^7$ are the present members. The group key $k$ and two node keys $k_{78}$ and $k_{58}$ are changed to $k'$, $k'_{78}$, and $k'_{58}$, respectively.

The key distribution process for each user join can be listed as:

1. Server authenticates the interested user $u^8$ and allocates it an empty place in the group tree. Server also provides individual key $k_8$ to the new user.

2. All the keys which $u^8$ needs in order to communicate with group members, i.e., $k'$, $k'_{78}$, and $k'_{58}$ are sent by unicast to $u^8$ encrypted with its individual key $k_8$.

3. $k'_{78}$ is shared by unicast to the sibling of new user $u^8$, $u^7$, after encrypting with its individual key $k_7$.

4. $k'_{58}$ is transmitted by multicast to $u^7$ and $(u^5, u^6)$ encrypted with $k'_{78}$ and $k_{56}$, respectively.

5. New group key $k'$ is sent by multicast to $(u^5, u^6, u^7)$ and $(u^1, u^2, u^3, u^4)$ after being encrypted with $k'_{58}$ and $k_{14}$, respectively.

### 2.1.3 On User Leave

On eviction of user, middle node keys must be changed to preserve communication secrecy in the group.

Suppose user $u^8$ leaves the group. Here, the group key $k$ and node key $k_{58}$ must be changed. Their new values can be represented as $k'$ and $k'_{57}$, respectively.

1. User $u^8$ node is deleted from the key tree at first.

2. User $u^8$ sibling's node, $u^7$, moves to its parent's node.

3. $k_{58}$ and $k$ are updated to new values, $k'$ and $k'_{57}$, respectively. These new values are then sent to users $u^7$, $(u^5, u^6)$, and $(u^1, u^2, u^3, u^4)$ by encrypting them with keys $k_7$, $k_{56}$, and $k_{14}$, respectively.

## 2.2 One-Way Function Key Tree

OFT key management scheme decreases server-level computation, as computation load is distributed between the server and group members. Rekeying overhead for OFT is $hk + h$.

Figure 2-2 shows OFT key management scheme, where $f(a,b)$ is a mixing function and $g(.)$ a one-way hash function. The value of $g(.)$ is called blinded node key.

Details of these functions are as follows.

- **One-way function,** $g(\cdot)$**:** The keys are passed through a strong one-way function to hide the contents of the original key. These "blinded" keys can be shared to corresponding users without any security concerns.

- **Joining function,** $f(a,b)$**:** This function concatenates or combines two entities, $a$ and $b$.

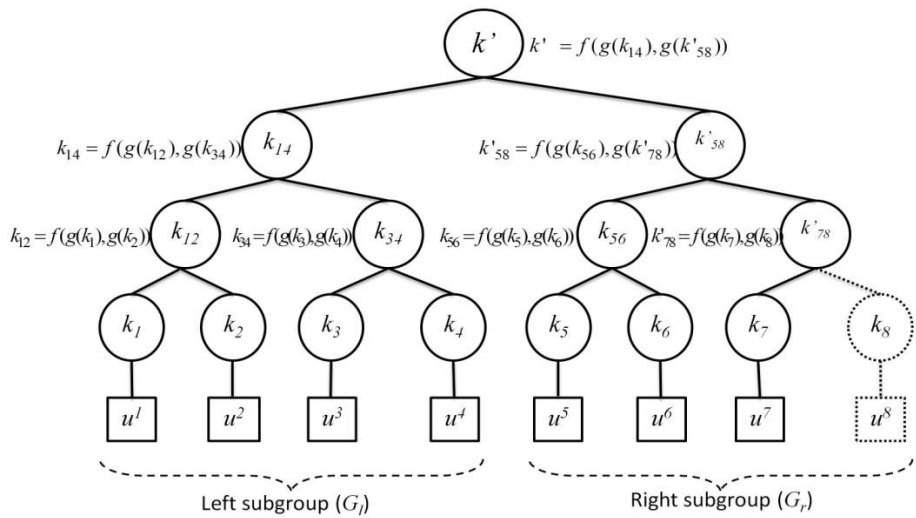

Figure 2-2 OFT key tree of height $h=3$.

Members can calculate desired key $k_i$ by following formula

$$k_i = f(g(k_{left(i)}), g(k_{right(i)}))$$

where $k_{left(i)}$ and $k_{right(i)}$ denote left and right children node keys of the node key $k_i$.

### 2.2.1 OFT Structure

In OFT, keys are dependent on each other. Group members have knowledge of certain blinded node keys, which enables them to generate new keys on every user join and leave. All members know their sibling's blinded node key as well as their ancestors' sibling's blinded node keys. For example, user $u^1$ also stores $g(k_2)$, $g(k_{34})$, and $g(k_{58})$ along with its individual key $k_1$. Now, members can compute their node keys and more importantly, the group key by using these known values. Left subgroup with users $(u^1, u^2, u^3, u^4)$ performs the following operations to find their node keys as

$$k_{12} = f(g(k_1), g(k_2))$$
$$k_{34} = f(g(k_3), g(k_4))$$
$$k_{14} = f(g(k_{12}), g(k_{34})).$$

Blinded subgroup node key $k_{14}$, which is left child node of group key, is shared with other subgroup. Users $(u^5, u^6, u^7, u^8)$ can compute key for right child node of group key, which is then shared with neighboring left subgroup as

$$k_{56} = f(g(k_5), g(k_6))$$
$$k_{78} = f(g(k_7), g(k_8))$$
$$k_{58} = f(g(k_{56}), g(k_{78})).$$

Group key $k$ can be generated by all members of the group as

$$k = f(g(k_{14}), g(k_{58})).$$

Here, server as well as all group users participates to compute the group key.

### 2.2.2 On User Join

Figure 2-2 shows the case of use joining, where a new user has just entered the group. When a user, say $u^8$, joins the group, he will receive his sibling's blinded node key $g(k_7)$, and his ancestors' sibling blinded node keys, $g(k_{56})$ and $g(k_{14})$. Server unicasts these keys to $u^8$ after encrypting them with individual key of $u^8$, $k_8$. On the other hand, $u^8$ calculates $k'_{78}$, $k'_{58}$, and $k'$ by using the following formulae

$$k'_{78} = f(g(k_7), g(k_8))$$
$$k'_{58} = f(g(k_{56}), g(k'_{78}))$$
$$k' = f(g(k_{14}), g(k'_{58})).$$

The blinded values of the calculated keys are encrypted with their sibling keys and advertised to existing group members by multicast as follows:

$$server \xrightarrow{multicast} \begin{cases} u^1 \sim u^4 : \{g(k'_{58})\}k_{14} \\ u^5 \text{ and } u^6 : \{g(k'_{78})\}k_{56} \\ u^7 : \{g(k_8)\}k_7 \end{cases}$$

where $\{g(\cdot)\}k_i$ denotes the encryption of value $g(\cdot)$ by key $k_i$.

### 2.2.3 On User Leave

On every user leave, group key must be changed to preserve forward and backward secrecy. The sibling of the evicted user is assigned with new key and it moves up to their parent's node. The group key alters because of these steps.

Suppose user $u^8$ leaves the group. Group key $k'$ and node key $k'_{58}$ are computed by members of the group by using blinded node keys given as

$$k'_{58} = f(g(k_{56}), g(k'_{7}))$$
$$k' = f(g(k_{14}), g(k'_{58})).$$

Updated right subgroup key $k'_{58}$ is shared with complementary left subgroup, whose members can calculate the new group key.

By using this procedure, the need of multicasting all updated keys is reduced, as the users can compute necessary keys themselves. Only few blinded node keys are sent to particular users and subgroups. OFT reduces the required broadcasts to nearly half as compared to LKH scheme.

## 2.3 Collusion Attacks on OFT

### 2.3.1 Horng's Attack

Horng [6] shows that OFT scheme is susceptible to collusion attacks, where leaving and joining users can collude their information of group keys to find older or newer group keys as shown in Figure 2-3. This weakens the security notions of forward and backward secrecy.

Here we present an example of such an attack on OFT.

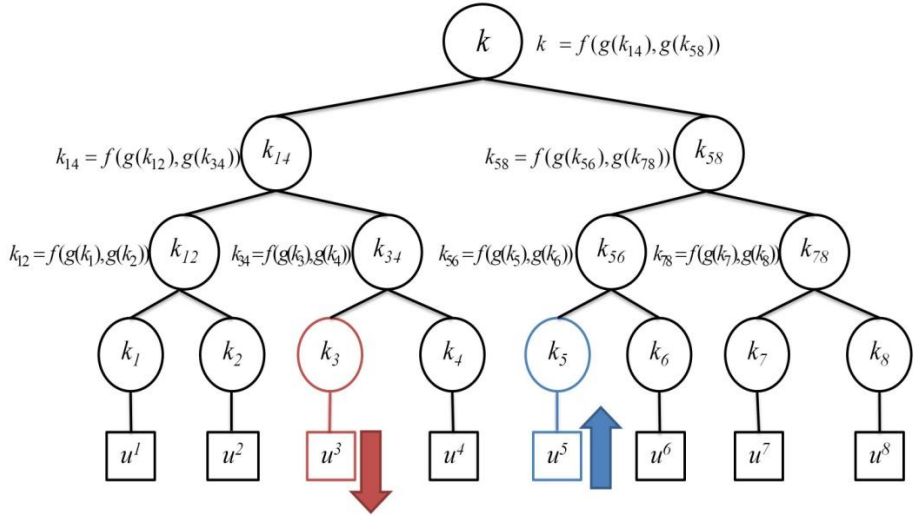

Figure 2-3 User $u^3$ is leaving the group and user $u^5$ joins the group.

Suppose the initial group key to be $k_{t0}$ given as

$k_{t0} = f(g(k_{14}), g(k_{58}))$.

In Figure 2-3, user $u^3$ leaves the group. This causes the group key to be changed. New group key $k_{t1}$ will be

$k_{t1} = f(g(k'_{14}), g(k_{58}))$.

If there is no other key updating operation and user $u^5$ joins the group, then new group key $k_{t2}$ will be

$k_{t2} = f(g(k'_{14}), g(k'_{58}))$.

User $u^3$ knows the value $g(k_{58})$ and user $u^5$ has knowledge of $g(k'_{14})$. Both users can collude their information to form $k_{t1}$ as $k_{t1} = f(g(k'_{14}), g(k_{58}))$. OFT scheme is unsuccessful to provide forward secrecy against $u^3$ and backward secrecy against $u^5$.

### 2.3.2 Ku and Chen's Attack

Ku and Chen [10] present some other cases of collusion attacks on OFT. We present these conditions here.

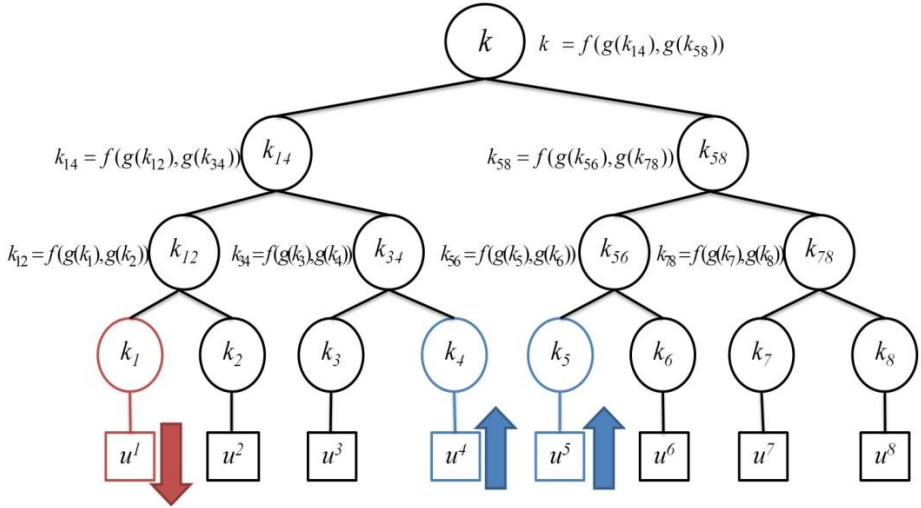

Figure 2-4 User $u^1$ leaves the group and users $u^4$ and $u^5$ join the group.

Consider Figure 2-4 for these attacks. Alice is represented by user $u^1$ in the group; Bob is associated with user $u^4$ in the key tree; Candy is related to user $u^5$. Also, time intervals follow the relation $t_3 > t_2 > t_1$.

If Alice is evicted at state $t_1$ and Bob is added to the same subgroup at time $t_2$, they can collude to get the value of $G_l$ subgroup key between time intervals $t_1$ and $t_2$, as observed in Horng's attack. As both members also know the blinded node key of subgroup $G_r$, they can easily compute group key between time intervals $t_1$ and $t_2$. Also, consider that Alice leaves the group at time $t_1$.

After this, Bob joins the group at time $t_2$, whereas Candy joins the group at time $t_3$ at locations shown in Figure 2-4.

Alice and Bob are associated with $G_l$, whereas Candy belongs to $G_r$. Here, Alice knows the blinded node key of subgroup $G_r$ between times $t_1$ and $t_3$. Candy knows the blinded node key of subgroup $G_l$ between time $t_2$ and $t_3$. Alice and Candy can share their information about blinded node keys of subgroups to compute group key between time $t_2$ and $t_3$.

In both of these cases, successful collusion attacks occur, which compromise the security of group communication. OFT scheme has security vulnerabilities, which must be addressed.

## 2.4 Security Requirements

Now we list some security requirements which must be fulfilled by all key management schemes in order to ensure secure communication in dynamic multicast systems.

1. **Group key secrecy:** Any passive adversary is unable in any way to compute previous or existing group key. This also implies that adversary is also unable to find any changed node key in the group. Mathematical operations and random numbers involved in rekeying must be cryptographically strong.

2. **Forward key secrecy**: Passive adversaries or former members of the group, who may know any subset of older group keys, cannot find any new group key.

3. **Backward key secrecy**: Passive adversaries or present members of the group, who may know any subset of group keys, are unable to discover any previously used group key.

4. **Key independence**: Passive adversaries or former and present members of the group, who may know any subset of group keys, are unable to discover any other group key.

5. **Reuse of known node keys**: Evicted members must not discover any new information that is flowing within the group. Sometimes evicted users can use their prior knowledge of node keys to decrypt any future transmission. All node keys known to a leaving member must be changed during rekeying process.

6. **OFT group key segments**: Group key in OFT scheme is combination of blinded node keys of its two children. These two children nodes represent left and right subgroup node keys. For each user leave, both segments of the group key must be changed. This step prevents occurring of any collusion attacks.

# 3 New Secure OFT Scheme

In this chapter, improvement in the OFT schemes will be introduced, that is, OFT scheme is vulnerable to collusion attacks and thus this scheme needs extra steps to make it reliable enough for proper functionalities.
We propose new OFT scheme with more security and lesser costs.

## 3.1 Introduction

In OFT scheme, all members are given the blinded node keys of their siblings and their ancestors' siblings. They can then calculate the desired node and group keys by using these blinded node keys. In this way, a part of computation load is transferred to the users from the server. In OFT, rekeying overhead decreases as a result of combined computations by the server and members.

First we define the levels and locations of the members in the group. Figure 3-1 shows how we name each user based on their level and location within level in the group. This will help us in introducing the scheme which provides better key management performance along with ensuring better security against the collusion attacks.

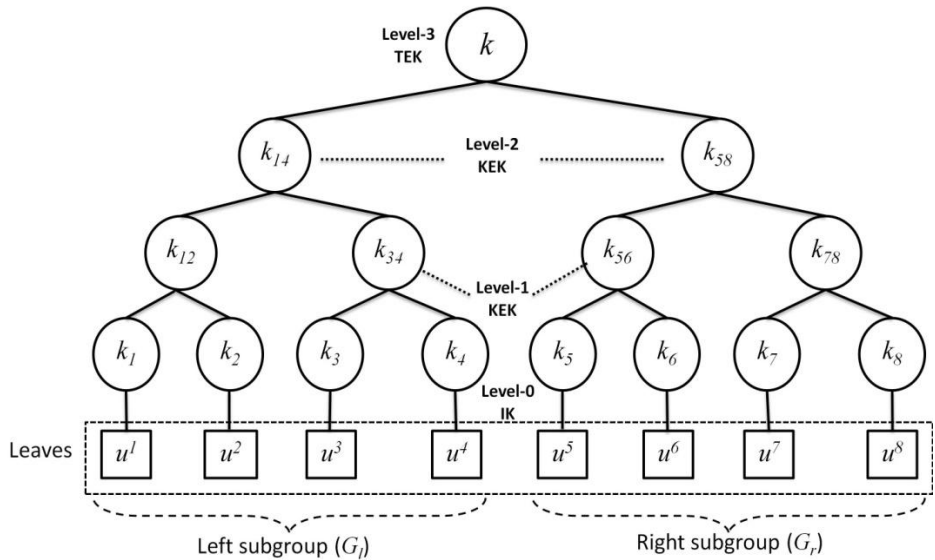

Figure 3-1 Key tree with levels.

In the above figure,

- **TEK:** traffic encryption key (group key), used for communicating with all the users in the group (highest level node key)
- **KEK:** key encryption keys, also called as sub-group keys. They are used for encrypting the group key for its transfer (intermediate level node keys).
- **IK:** Individual keys of users (lowest level node keys)
- **Height of tree ($h$):** Number of levels in the tree. For example, the tree shown in Figure 3-1 has height of 3.

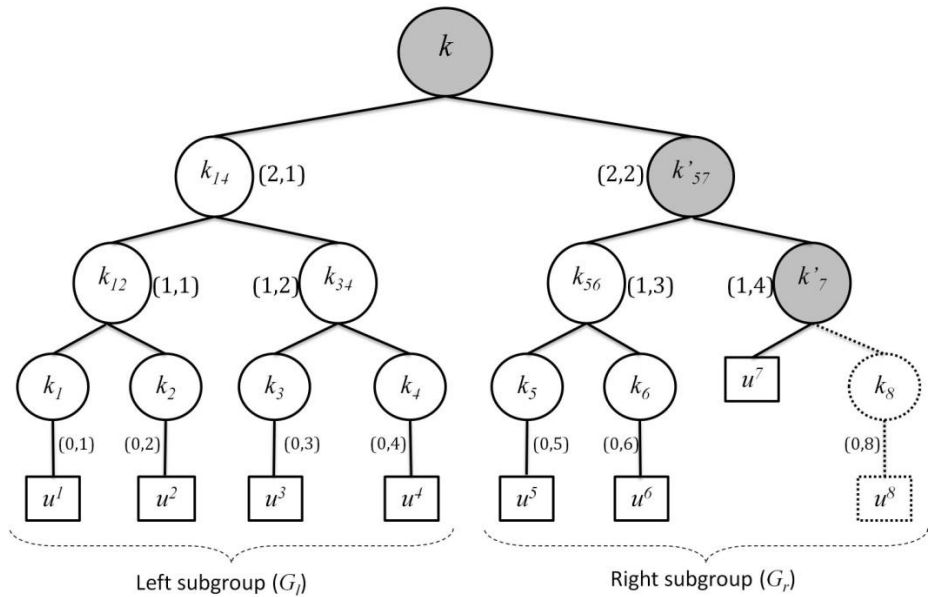

Figure 3-2 Key tree with location indices.

We provide all the users with location indices which give information about their level and location in the group. Users must have $\log_2 n$ blinded node keys, which are the blinded node keys of their siblings and their ancestors' siblings, to communicate within group and update the keys whenever needed. These blinded node keys are kept with their indices, in the user's memory as shown in Figure 3-2. This helps in maintaining and updating these keys whenever there is some change in the group.

For example, the keys with the user $u^1$ can be written as shown in Table 3.1.

Table 3.1 The keys with user $u^1$ (0,1)

| Indices | Key stored |
|---------|------------|
| (0,2)   | $g(k_2)$   |
| (1,2)   | $g(k_{34})$ |
| (2,2)   | $g(k_{58})$ |

Similarly, other users have the blinded node keys stored with the same pattern.

Our scheme performs only at user eviction. On user join, it follows the scheme of original OFT.

## 3.2 On User Leave

In case of eviction, the sibling of the removed user takes the place of their parent. Also its sibling key changes, which causes change in the node key and the group key. Figure 3-2 shows eviction of $u^8$ after which $u^7$ promotes to a higher level. Shaded nodes are the ones to be altered.

### 3.2.1 Key Requirements

Node keys are dependent in OFT scheme and users contribute in generation of node keys. Users have blinded node keys of their siblings and parent's siblings, which are used to efficiently generate new keys without much intervention from server. Group members update the keys present with them and generate new node keys and the group key. New keys can be generated as explained in the following protocol.

### 3.2.2 Algorithm on User Eviction

We present the algorithm which is followed by our scheme at user eviction.

1. Server removes the node of evicted member from the key tree and promotes its sibling, if any, to higher level which was the level of its parent before eviction.
2. Server provides new node key to evicted member's sibling. It also shares the changed blinded node keys with appropriate neighboring members.
3. All members of the affected subgroup can compute new subgroup node key.
4. Server shares this new subgroup node key in blinded form with neighboring unaffected subgroup members.
5. For the unaffected subgroup, server generates and shares a random number with its members.
6. Members update stored node keys by XORing them with the provided random number, after which the resulting values are passed through one-way function to generate new node keys.
7. All members of unaffected subgroup can calculate new subgroup node key by utilizing one-way and combining functions on blinded node key.
8. This new subgroup node key is then shared with the neighboring (unaffected) subgroup.
9. All members of the group can use blinded node keys of affected

and unaffected subgroup to calculate new group key.

### 3.2.3 An Example of User Eviction

User $u^8$ leaves the subgroup with users $u^5, u^6, u^7$, and we define this subgroup as right, $G_r$. The other subgroup is defined as left subgroup, $G_l$. Group key is combination function of blinded node keys of subgroups $G_r$ and $G_l$. Our protocol changes both of these subgroup keys in order to prevent collusion attacks.

1. After removal of $u^8$, $k_7$ changes and this affects the node keys, which changes to $k'_{57}$ and $k'$. The new blinded node key $g(k'_7)$ is shared with the neighboring subgroup $(u^5, u^6)$ by encrypting with their node key $k_{56}$ as

    $u^5$ and $u^6 : \{g(k'_7)\}k_{56}$.

2. Subgroup key $k'_{57}$ can be formed by

    $k'_{57} = f(g(k'_7), g(k_{56}))$.

3. Right subgroup key is shared with the other subgroup as

    $G_s : \{g(k'_{57})\}k_{14}$.

4. Level-2 blinded subgroup node key is transmitted to the left subgroup members, encrypted with their blinded subgroup node key as

    $G_s : \{g(k'_{58})\}k_{14}$.

5. As for the left subgroup, server generates a random number $r_n$, which is shared between all the present users of the subgroup as

    $G_s : \{r_n\}k_{14}$.

6. Users can change the blinded node keys with them, so as to alter the overall left blinded subgroup node keys. Already available blinded node keys can be changed like as shown in Table 3.2.

Table 3.2 Blinded node key change operations for user $u^1(0,1)$

| Original blinded node keys | New blinded node keys |
|---|---|
| $k_1, g(k_2)$ | $k_1, g(k_2)$ |
| $g(k_{12})$ | $g(k'_{12}) \rightarrow g(g(k_{12}) \oplus r_n)$ |
| $g(k_{34})$ | $g(k'_{34}) \rightarrow g(g(k_{34}) \oplus r_n)$ |
| $g(k_{14})$ | $g(k'_{14}) \rightarrow g(g(k_{14}) \oplus r_n)$ |
| $g(k_{58})$ | $g(k'_{57})$ |

7. New node key can be formed by all left subgroup users as shown in Table 3.2.

    $g(k'_{14}) \rightarrow g(g(k_{14}) \oplus r_n)$

8. This level-2 blinded subgroup node key is transmitted to the right subgroup members, encrypted with their blinded subgroup node key as

    $G_p : \{g(k'_{14})\}k'_{58}$.

9. New group key $k$ can be computed by all members of the group by using

    $k = f(g(k'_{14}), g(k'_{58}))$.

## 3.3 Simulation and Results

Results of our proposed scheme and the conventional key management schemes are shown in this section.

Table 3.3 shows security properties of some of the known schemes, and shows that our proposed scheme has got better security strength.

Table 3.3 Security of key management schemes

| Schemes | Secrecy | | Secure against collusion attacks |
|---|---|---|---|
| | Forward | Backward | |
| Simple | Y | Y | Y |
| GKMP [4] | N | Y | Y |
| LKH [9] | Y | Y | Y |
| OFT [7] | N | N | N |
| Ku and Chen [10] | Y | Y | Y |
| Xu *et al.* [11] | Y | Y | Y |
| Proposed sol. | Y | Y | Y |

Tables 3.4 and 3.5 show performance of various schemes. They show that our scheme has less broadcast costs for user leave, as compared to scheme by Ku and Chen [10]. Our scheme performs even more efficiently than OFT for large group sizes.

Table 3.4 Performance comparison of key management schemes

| Schemes | Message | | |
|---|---|---|---|
| | Join | | Leave |
| | multicast | unicast | |
| Simple | $nK$ | $K$ | $nK$ |
| GKMP | $2K$ | $2K$ | - |
| LKH | $2\log_2(n)$ | $\log_2(n)$ | $2\log_2(n)$ |
| OFT | $\log_2(n)$ | $\log_2(n)$ | $\log_2(n)$ |
| Ku and Chen | $\log_2(n)$ | $\log_2(n)$ | $(\log_2(n))^2 + \log_2(n)$ |
| Proposed sol. | $\log_2(n)$ | $\log_2(n)$ | 5 |

Table 3.5 Encryption costs for different schemes

| Schemes | Join | Leave |
|---------|------|-------|
| LKH | $3\log_2(n)$ | $2\log_2(n)$ |
| OFT | $2\log_2(n)+2$ | $\log_2(n)$ |
| Ku and Chen | $2\log_2(n)+2$ | $(\log_2(n))^2+\log_2(n)$ |
| Proposed sol. | $2\log_2(n)+2$ | 5 |

## 3.4 Security Analysis of Proposed Scheme

Possibility of collusion attacks in OFT scheme arises because OFT keeps the blinded node keys known to the evicted members intact. Evicted members can then share blinded node keys available with them with other users in order to get the group key information which they are not intended to know.

The proposed scheme, on the other hand, amends this vulnerability, that is, blinded subgroup node keys known to the evicted members are changed. In the proposed schemes, the former members are not able to reconstruct group keys by exercising any attacks explained in this paper.

An example of a typical case of member eviction is given here. Alice evicts the subgroup $G_l$, after which Bob and Candy join the subgroup $G_l$ or $G_r$. Alice knows the initial group key

$k_{G(s_0)} = f(g(k_{G_l}), g(k_{G_r}))$.

Group keys at eviction and joining for different cases are given below.

**Case 1:**

Consider the case when both Bob and Candy join same subgroup, $G_r$.

Then, we can write group key $k_G$ at different steps as follows.

**Step 1:** *Alice evicts* $G_l$

$k_{G(s_1)} = f(g(k'_{G_l}), g(k'_{G_r}))$.

**Step 2:** *Bob joins* $G_r$

$k_{G(s_2)} = f(g(k'_{G_l}), g(k''_{G_r}))$.

**Step 3:** *Candy joins* $G_r$

$k_{G(s_3)} = f(g(k'_{G_l}), g(k'''_{G_r}))$.

As obvious in this case, Alice is not able to collude with any present member in order to find illegitimate group keys.

**Case 2:**

Consider the case when Bob joins subgroup $G_l$, whereas Candy joins the subgroup $G_r$.

Then, we can write group key $k_G$ at different steps as follows.

**Step 1:** *Alice evicts* $G_l$

$k_{G(s_1)} = f(g(k'_{G_l}), g(k'_{G_r}))$.

**Step 2:** *Bob joins* $G_l$

$k_{G(s_2)} = f(g(k''_{G_l}), g(k'_{G_r}))$.

**Step 3:** *Candy joins* $G_r$

$$k_{G(s_3)} = f(g(k''_{G_l}), g(k''_{G_r})).$$

As seen in this case also, Alice is not able to collude with present members in order to find illegitimate group keys.

Thus, the proposed scheme prevents any collusion attacks.

# 4 Multicast Scheme Based on LKH

In this chapter, we will provide a new scheme based on LKH. Our proposed scheme is more efficient than the original LKH scheme in terms of communication overheads needed at rekeying.

## 4.1 System Design

### 4.1.1 Design Principles

Firstly, we outline some basic principles, which our proposed scheme will comply. We use binary key tree, which tends to balance itself in order to maintain the symmetry. Interested hosts can join the group through a process which includes sending request to the server and passing the authentication by the server.

Server provides an empty place on the group to the new users. Server also provides the new member with all the necessary keys through the key management protocol. Similarly, present members can leave the group by sending request to the server, who in return eradicates the user and its corresponding node from the key tree. Sibling of the leaving node moves to the position of their parent node.

### 4.1.2 Detailed Outline

On each join or leave, keys in the path from that location to server need to be changed. This ensures the backward and forward secrecy requirements of secure group communication. Server changes $\log_2 n$ keys for each join, and $\log_2 n - 1$ keys for each user leave.

The keys, which intend to be changed, affects $2^l$ users, where $l$ is the level of the key. This calls for an efficient protocol, capable of sharing new keys among all members of the group.

Figure 4-1 shows a binary key tree with $l = 3$. On a user join, as shown in the figure, $\log_2 n$ keys, namely, $k$, $k_{58}$, and $k_{78}$ are affected. Change of $k_{78}$ will affect two users $u^7$ and $u^8$. Similarly, changing subgroup key $k_{58}$ affects four users $u^5 \sim u^8$.

Each group member needs to change its group key $k$, on every user join and leave.

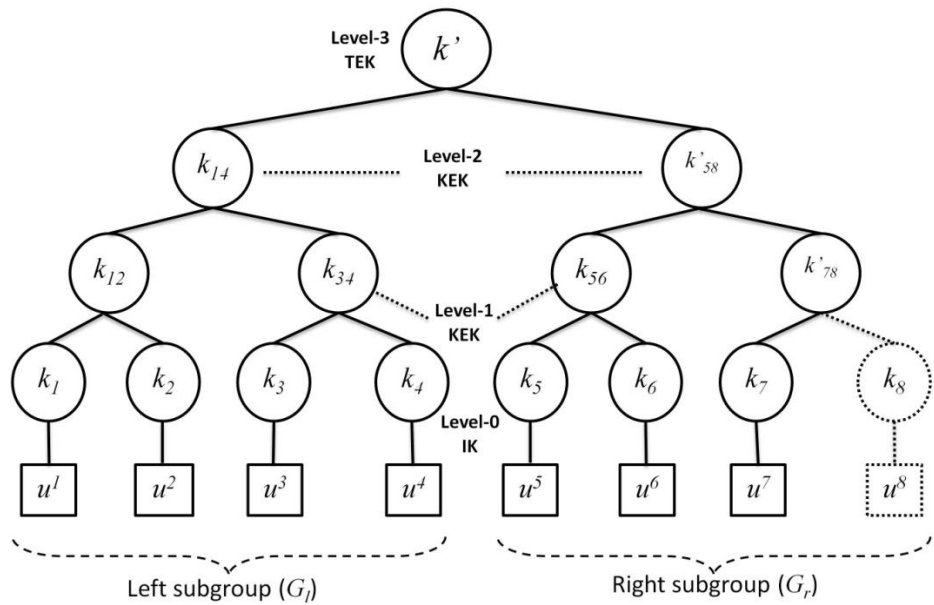

Figure 4-1 Binary key tree on a user join.

Our proposed scheme possesses the following distinctive features while distributing keys for right subgroup.

- Our scheme follows bottom-to-top approach, where bottom node keys are firstly distributed to the desired members, moving upwards.
- Higher level node keys are encrypted with lower level ones, and multicast to subsequent subgroups. For example, subgroup key at level-1 will be distributed by encrypting it with individual keys at level-0. After that, a subgroup key at level-2 is encrypted with level-1 key before multicast, and so on.
- Instead of wasting resources by sending all keys to only one user, our proposed scheme moves in a step-wise manner thus providing all essential keys to members.

## 4.2 On User Join

After the interested host is successfully authenticated, server allocates the host an empty location in the group. Group key tree is renewed, according to following protocol. Example protocol for height $h = 3$ is described below, which will be generalized afterwards. Figure 4-1 refers to the key tree for join case.

$u^8$ joins the group, forming a subgroup with $u^7$. The shaded keys in the figure are changed to new ones by the server.

### 4.2.1 Key Requirements

Depending on their location in the group, members require different keys to update the essential keys. Desired keys by user can be outlined as

$$u^1 \sim u^8 : k'$$
$$u^5, u^6, u^7, u^8 : k'_{58}$$
$$u^7 \text{ and } u^8 : k'_{78}.$$

$k'$, $k'_{58}$, and $k'_{78}$ are new group key and subgroup keys, respectively.

### 4.2.2 Protocol for User Join

To share keys among the members of the group, they are encrypted by individual or subgroup keys and sent through unicast or multicast to respective members.

$\{k\}k_{ij}$ represents encryption of the group key $k$ by any subgroup key $k_{ij}$. The same notation is used in describing the schemes.

The protocol for key management on user join is given below, where the keys are being transmitted by the server to various locations.

1. Server encrypts new group key $k'$ with subgroup key $k_{14}$ and multicasts it to left subgroup as

    $u^1 \sim u^4 : \{k'\}k_{14}.$

2. For right subgroup, key distribution starts at the bottom where the bottom-most node key $k'_{78}$ is encrypted with individual keys of both users $u^7$ and $u^8$, before being unicast to them as

    $u^7 : \{k'_{78}\}k_7$
    $u^8 : \{k'_{78}\}k_8.$

3. Level-2 node key $k'_{58}$ is encrypted with level-1 node keys $k_{56}$ and $k'_{78}$, and shared as

$u^5$ and $u^6 : \{k'_{58}\}k_{56}$

$u^7$ and $u^8 : \{k'_{58}\}k'_{78}$.

4. Right subgroup gets its new group key $k'$ by encrypting and multicasting it as

$u^4 \sim u^8 : \{k'\}k'_{58}$.

## 4.3 On User Leave

User leave makes an empty slot in the binary balanced key tree. Sibling of the leaving user gets promoted to the position of its parent's node.

Figure 4-2 shows user $u^8$ leaving the tree, after which user $u^7$ resides on level-1. Shaded nodes show the compromised keys that will be changed by the server.

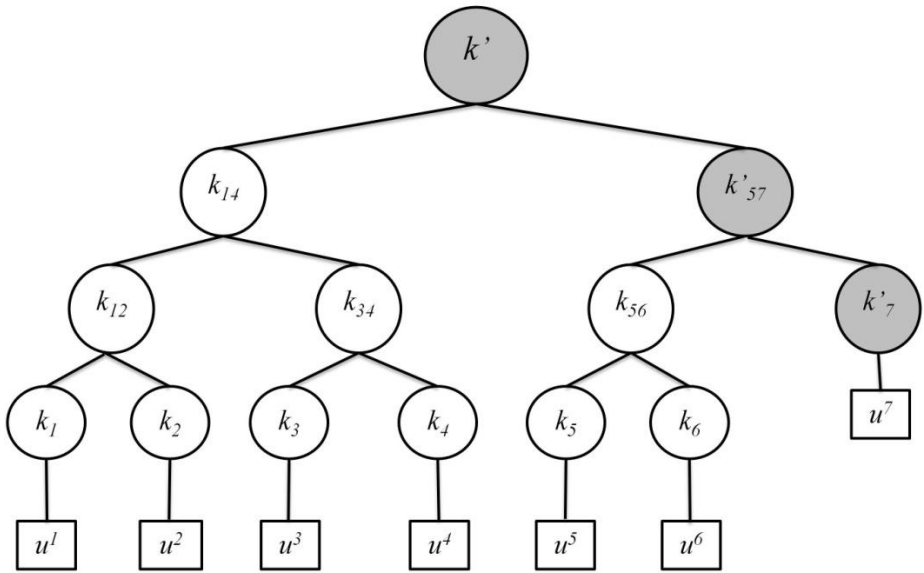

Figure 4-2 Binary key tree on a user leave.

### 4.3.1 Key Requirements

For each user leave, server generates and shares $\log_2 n - 1$ keys to remaining members. The demand for keys differs with the member's location. Keys needed for update can be outlined as

$$u^1 \sim u^7 : k'$$
$$u^5, u^6, u^7 : k'_{57}$$

where $k'$ and $k'_{57}$ are new group key and subgroup key, respectively.

### 4.3.2 Protocol for User Leave

Less number of keys is shared on each user leave, contrary to the number of keys distributed for each user join.

The protocol for key management on each user leave is described below, where the keys are being transmitted by the server to various locations encrypted by either individual or subgroup keys.

1. Server encrypts new group key $k'$ with subgroup key $k_{14}$ and multicasts it to left subgroup as

    $$u^1 \sim u^4 : \{k'\}k_{14}.$$

2. As for right subgroup, key distribution starts at the bottom. Level-2 node key $k'_{57}$ is encrypted with node key $k_{56}$ and individual key of user $u^7$, $k_7$, before being multicast and unicast, respectively.

    $$u^5 \text{ and } u^6 : \{k'_{57}\}k_{56}$$
    $$u^7 : \{k'_{57}\}k_7$$

3. Right subgroup gets its new group key $k'$ by encrypting and multicasting it as

$$u^4 \sim u^7 : \{k'\}k'_{57}.$$

## 4.4 Simulation and Results

Performance of our proposed scheme as compared to LKH scheme is shown in this section.

### 4.4.1 Performance Comparison

Tables 4.1 and 4.2 show performance of various schemes. They show that our scheme has less broadcast costs for user leave and join as compared to other key management schemes. Our scheme performs more efficiently than LKH and the communication overhead of our scheme is less than that of LKH.

Table 4.1 Performance comparison of hierarchical schemes

| Schemes | Message | | Leave |
|---|---|---|---|
| | Join | | |
| | Multicast | Unicast | |
| Simple | $nK$ | $K$ | $nK$ |
| GKMP | $2K$ | $2K$ | - |
| LKH | $2\log_2(n)$ | $\log_2(n)$ | $2\log_2(n)$ |
| Our solution | $\log_2(n)$ | $\log_2(n)-1$ | $\log_2(n)+1$ |

Encryption cost of our proposed scheme is also less than the original LKH scheme.

Table 4.2 Encryption costs for different schemes

| Schemes | Join | Leave |
|---|---|---|
| LKH | $3\log_2(n)$ | $2\log_2(n)$ |
| Our solution | $2\log_2(n)$ | $2\log_2(n)-2$ |

### 4.4.2 Empirical Analysis

Following table shows broadcast costs at user join and leave for LKH and our scheme. The improvement in results can be analyzed from the table.

Table 4.3 shows that our scheme has less broadcast costs for user join and leave, as compared to LKH. Thus, our scheme is more efficient than LKH in terms of costs.

Table 4.3 Broadcast costs of schemes

| Schemes<br>Height | LKH | | Our solution | |
|---|---|---|---|---|
| | Join | Leave | Join | Leave |
| 10 | 30 | 20 | 20 | 18 |
| 12 | 36 | 24 | 24 | 22 |
| 13 | 39 | 26 | 26 | 24 |
| 14 | 42 | 28 | 28 | 26 |
| 15 | 45 | 30 | 30 | 28 |
| 16 | 48 | 32 | 32 | 30 |
| 17 | 51 | 34 | 34 | 32 |
| 18 | 54 | 36 | 36 | 34 |

Table 4.4 compares our proposed scheme at user join and leave with original LKH scheme. It is clear that even for large groups, our proposed scheme gives better overhead costs.

Table 4.4 Comparison of our solution

| Height<br>Our sol./LKH | 10 | 12 | 13 | 14 | 15 | 16 | 17 | 18 |
|---|---|---|---|---|---|---|---|---|
| Join | 0.67 | 0.67 | 0.67 | 0.67 | 0.67 | 0.67 | 0.67 | 0.67 |
| Leave | 0.9 | 0.917 | 0.923 | 0.928 | 0.933 | 0.9375 | 0.941 | 0.944 |

Following Figures 4-3 and 4-4 show the performance comparison of our proposed scheme with LKH for different heights of key tree.

The following figures also show that our proposed scheme performs better with different number of users.

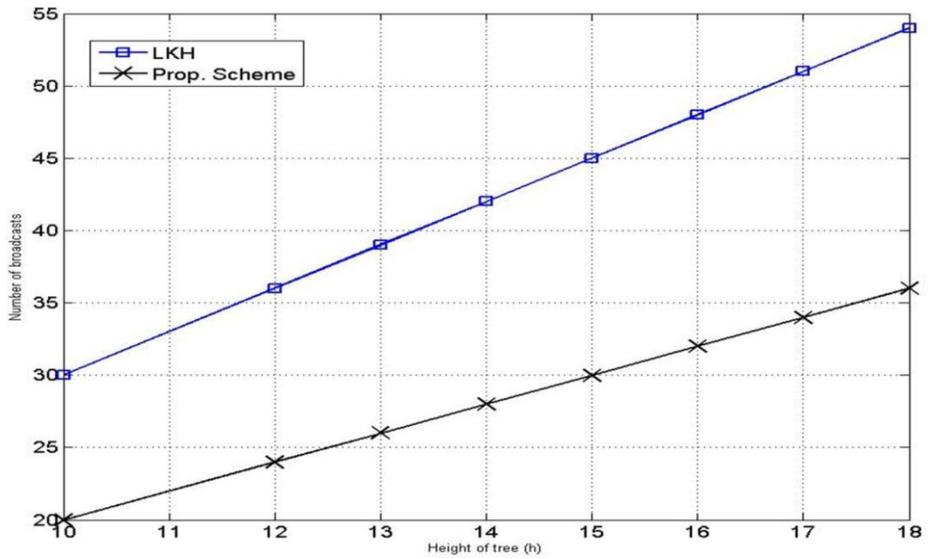

Figure 4-3 Comparison between the proposed scheme and LKH on user join.

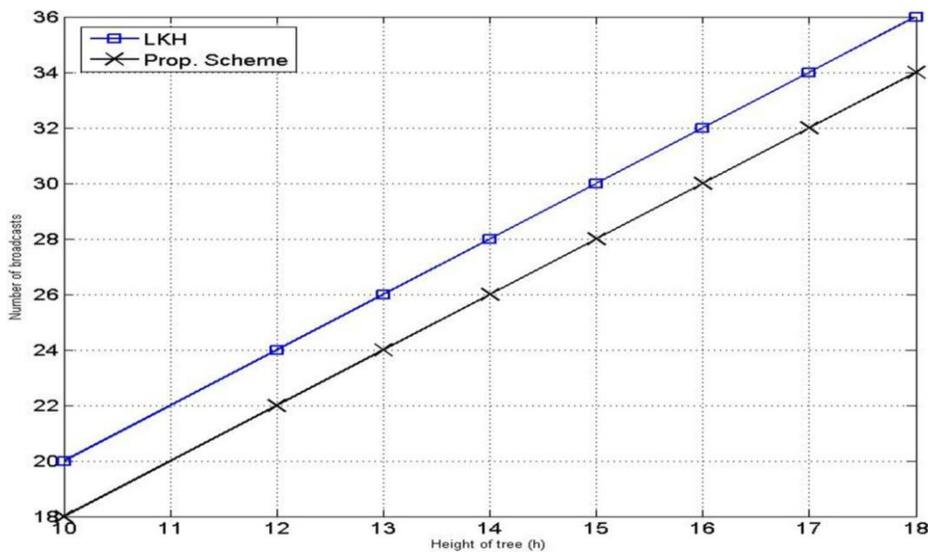

Figure 4-4 Comparison between the proposed scheme and LKH on user leave.

# 5 Conclusion

Key management in dynamic groups, where users can leave or join at their ease is an important part of secure communication. Different strategies have been proposed during last decade that aim to either improve the security or the performance of key management schemes. Decreasing the encryption and transmission overheads has also been a major concern for such schemes.

In this thesis, we proposed two schemes based on different architectures. One of the schemes improves the security of OFT scheme. We showed the resilience of proposed scheme by analyzing different cases. The other proposed scheme improves the performance of independent key hierarchy system (LKH).

Both proposed schemes provide better broadcast and transmission costs than previously published schemes.